\documentclass[preprint,showpacs,preprintnumbers,amsmath,amssymb,nofootinbib]{revtex4}
\usepackage{graphicx}
\usepackage{subfigure}
\usepackage{dcolumn}
\usepackage{bm}

\def \be {\begin{equation}}
\def \ee {\end{equation}}
\def \bea{\begin{eqnarray}}
\def \eea{\end{eqnarray}}

\def \prd{{\it Phys. Rev. D, }}
\def \prl{{\it Phys. Rev. Lett., }}
\def \apj{{\it Ap. J., }}
\def \GRG{{\it Gen. Relativ. Gravit., }}

\begin{document}
\title{LISA Sensitivities to Gravitational Waves from Relativistic
  Metric Theories of Gravity}

\author{Massimo Tinto}
\email{Massimo.Tinto@jpl.nasa.gov}
\affiliation{Jet Propulsion Laboratory, California Institute of
  Technology, Pasadena, CA 91109}
           
\author{M\'arcio Eduardo da Silva Alves}
\email{alvesmes@unifei.edu.br} 
\affiliation{Instituto de Ci\^ encias
  Exatas, Universidade Federal de Itajub\'a, 37500-903 Itajub\'a, MG,
  Brazil}

\date{\today}

\begin{abstract}
  The direct observation of gravitational waves will provide a unique
  tool for probing the dynamical properties of highly compact
  astrophysical objects, mapping ultra-relativistic regions of
  space-time, and testing Einstein's general theory of relativity.
  LISA (Laser Interferometer Space Antenna), a joint NASA-ESA mission
  to be launched in the next decade, will perform these scientific
  tasks by detecting and studying low-frequency cosmic gravitational
  waves through their influence on the phases of six modulated laser
  beams exchanged between three remote spacecraft.  By directly
  measuring the polarization components of the waves LISA will detect,
  we will be able to test Einstein's theory of relativity with good
  sensitivity.  Since a gravitational wave signal predicted by the
  most general relativistic metric theory of gravity accounts for {\it
    six} polarization modes (the usual two Einstein's tensor
  polarizations as well as two vector and two scalar wave components),
  we have derived the LISA Time-Delay Interferometric responses and
  estimated their sensitivities to vector- and scalar-type waves.  We
  find that (i) at frequencies larger then roughly the inverse of the
  one-way light time ($\approx 6 \times 10^{-2} $ Hz.)  LISA is more
  than ten times sensitive to scalar-longitudinal and vector signals
  than to tensor and scalar-transverse waves, and (ii) in the low part
  of its frequency band is equally sensitive to tensor and vector
  waves and somewhat less sensitive to scalar signals.
\end{abstract}

\maketitle

\section{Introduction}
\label{intro}

The direct detection of gravitational waves will represent one of the
greatest triumphs of experimental physics of this century, and provide
us with a new observational tool for obtaining better and deeper
understanding about their sources.

Several experimental efforts have been underway for years, both on the
ground and in space \cite{LIGO,VIRGO,GEO,TAMA,PULSAR,DOPPLER}, and
only recently kilometer-size ground-based interferometers have been
able to identify the most stringent upper-limits to date for the
amplitudes of the radiation expected from several classes of sources.
Although an unambiguous detection has not been declared with
present-generation instruments, next-generation Earth-based
interferometers and pulsar-timing experiments, as well as the LISA
(Laser Interferometric Space Antenna) mission \cite{PPA98} are
expected to achieve this goal.

LISA, jointly proposed to the National Aeronautics and Space
Administration (NASA) and the European Space Agency (ESA), is expected
to be flown sometimes in the next decade.  Its goal is to detect and
study gravitational waves (GW) in the millihertz frequency band. It
will use coherent laser beams exchanged between three identical
spacecraft forming a giant (almost) equilateral triangle of side $5
\times 10^6$ kilometers. By monitoring the relative frequency changes
of the light beams exchanged between the spacecraft, it will extract
the information about the gravitational waves it will observe at
unprecedented sensitivities \cite{PPA98}. The astrophysical sources
that LISA is expected to observe within its operational frequency band
($10^{-4} - 1$ Hz) will be very large in number, including galactic
and extra-galactic coalescing binary systems containing white dwarfs
and neutron stars, extra-galactic super-massive black-hole coalescing
binaries, and a stochastic gravitational wave background from the early
universe \cite{Thorne1987,SS2009}.

The first unambiguous detection of a gravitational wave signal,
whether performed on the ground or in space, will also allow us to
test Einstein's general theory of relativity (GR) by measuring the
polarization components of the detected signals
\cite{Nishizawa2009,Lee2008}. Among all the proposed relativistic
metric theories of gravity \cite{Will2006}, GR is the most
restrictive, allowing for only two of possible {\it six} different
polarizations \cite{Eardley1973}. By asserting that the spin-2
(``tensor'') polarizations are the only polarization components
observed, we would make a powerful proof of the validity of Einstein's
theory of relativity, while a clear observation of some other
polarization modes would disqualify it. Corroboration of polarization
measurements with estimates of the propagation speed of the observed
gravitational wave signal will provide further insight into the nature
of the observed radiation and result into the determination of the
mass of the graviton. It should be emphasized, however, that a
propagation-speed measurement alone consistent with a value equal to
the speed of light, would not automatically rule-out other
relativistic metric theories of gravity.  This is because waves with
helicity $s = 0$ can also propagate at light-speed \cite{Schutz2009}.
Gravitational waves with scalar polarization are predicted by the most
common generalizations of GR such as scalar-tensor theories. Besides
the classical example of Brans-Dicke theory \cite{Brans1961},
scalar-tensor theories result in the low-energy limit of string theory
(see, e.g., \cite{Casas1991}). Modifications of the Einstein-Hilbert
action, which consider generic functions of the Ricci scalar in the
Lagrangian (f(R) theories), also predict ``scalar'' gravitational
waves \cite{Alves2009}.  Vector modes, on the other hand, can appear
in the so-called ``quadratic gravity'' formulations \cite{Alves2009},
and in the context of theories in which the graviton has a finite mass
such as the Visser theory \cite{de Paula2004}.

LISA will not be able to distinguish the propagation speeds of scalar
(helicity $s = 0$) and vector (helicity $s = \pm 1$) polarizations
from the speed of light (a result following, as we shall see below,
from a combination of the existing stringent upper limits on the mass
of the graviton \cite{Talmadge1988,finn2002,deAraujo2007} and the LISA
observational bandwidth). However, it should be able to assess the
polarizations of the observed gravitational wave signals.  Since the
accuracy by which LISA will distinguish one polarization from another
will depend on their signal-to-noise ratios \cite{Helstrom1968}, in
this paper we estimate the LISA sensitivities to scalar- and vector-
polarized wave. 
 
The paper is organized as follows.  In Section (\ref{one-way}) we
derive the one-way Doppler response to a gravitational wave signal
characterized by six polarizations (2 ``tensor'' (helicity $s = \pm
2$), 2 ``vector'' (helicity $s = \pm 1$), and 2 ``scalar'' (helicity
$s = 0$)).  Since the resulting expression is equal {\it in form} to
the one previously derived by Estabrook and Wahlquist \cite{EW1975}
for tensor waves (i.e. for waves predicted by General Relativity), we
conclude that the responses of the various time-delay interferometric
(TDI) combinations are also identical {\it in form} to those
previously derived within the framework of GR \cite{AET1999}. Although
the derivation of the response function of a Michelson interferometer
to non-tensor polarization modes has been considered in previous
publications \cite{Nishizawa2009,Nishizawa2010}, the expression
presented there was correct in the so called
``long-wavelength-limit'', i.e. when the wavelength of the GW is much
larger than the size of the detector. Since LISA will be sensitive,
over most of its observational frequency band, to GWs of wavelength
shorter than its linear size, we have derived the expressions of the
LISA TDI responses to vector- and scalar- waves that are valid for any
wavelength. In Section (\ref{sensitivity}) we then compute the LISA
sensitivities to vector and scalar waves. We find that (i) at
frequencies larger then the inverse of the one-way light time
($\approx 6 \times 10^{-2}$ Hz) LISA is ten-times more sensitive to
scalar-longitudinal and vector signals than to tensor and
scalar-transverse waves, while (ii) it is equally sensitive to tensor
and vector waves, and somewhat less sensitive to scalar signals in the
low part of its frequency band.  Although both these results might
seem surprising at first, we show that they are consequence of the
physical and geometrical properties of the vector and scalar waves and
how they affect the frequency of the light beams exchanged by the
three LISA spacecraft. Finally in Section (\ref{Conclusions}) we
present a summary of the paper and our concluding remarks. Throughout
the paper we will be using natural units ( $c = G = h = 1$) except
where mentioned otherwise.

\section{Derivation of the One-Way Doppler Response}
\label{one-way}

In what follows we present the derivation of the ``one-way'' Doppler
response to a gravitational wave signal predicted by the most general
relativistic metric theory of gravity \cite{Eardley1973}.  Although
the result has already appeared in the literature \cite{Hellings1978},
our derivation results into an expression that is compact and
identical {\it in form} to that first obtained by Wahlquist
\cite{Wahlquist1987} in the case of plane gravitational waves
predicted by GR. This of course simplifies significantly the
derivation of the LISA TDI responses as they turn out to be equal {\it
  in form} to those given in \cite{AET1999}. As in
\cite{EW1975,Wahlquist1987}, our derivation is general and does not
rely on any assumptions about the size of the wavelength of the
radiation relative to the size of the detector.

Let us consider a space-time with the following metric
\begin{equation}
ds^2 = - dt^2 + (\delta_{ij} + h_{ij}(vt - z))dx^i dx^j,
~~|h_{ij}|<<1,
\label{GW metric}
\end{equation}
where the usual sum convention over repeated indices is assumed, Latin
indices go from $1$ to $3$, and a plane- gravitational wave has been
assumed, without loss of generality, to propagate along an arbitrary
$+z$ direction.  In Eq. (\ref{GW metric}) we have allowed the wave to
travel at a finite speed (group velocity) $v < 1$ to account for a
possible non-zero mass of the graviton.  Working in the context of GR,
it is well know that $h_{ij}$ has two degrees of freedom representing
gravitational waves (GWs) with helicity $s = \pm 2$, and $v = 1$. On
the other hand, alternative relativistic metric theories of gravity
allow for GWs with up to six degrees of freedom \cite{Eardley1973}.
Therefore, in the most general case, $h_{ij}$ can be represented in
terms of six components (with corresponding six metric amplitudes) in
the following form \cite{Eardley1973}
\begin{equation}\label{expansion}
h_{ij}(vt - z) = \sum_{r=1}^6 \epsilon_{ij}^{(r)}h_{(r)}(vt - z) \ ,
\end{equation}
where $\epsilon_{ij}^{(r)}$ are the six polarization tensors
associated with the six waveforms of the gravitational wave signal. If
we introduce a set of Cartesian orthogonal coordinates ($x, y, z$)
associated with the wave, in which ($x, y$) are in the plane of the
wave and $z$ is along the direction of propagation of the wave and
orthogonal to the ($x, y$) plane, the above six polarization tensors
assume the following matricial form \cite{Eardley1973}
\begin{equation}\label{polarization matrices}
   \begin{array}{cc}
      \left[\epsilon^{(1)}\right]_{ij} = \left(
        \begin{array}{ccc}
          0 & 0 & 0 \\
          0 & 0 & 0 \\
          0 & 0 & 1
        \end{array}
      \right) &

      \left[\epsilon^{(2)}\right]_{ij} = \left(
        \begin{array}{ccc}
          0 & 0 & 1 \\
          0 & 0 & 0 \\
          1 & 0 & 0
        \end{array}
      \right) \\
      \\

      \left[\epsilon^{(3)}\right]_{ij} = \left(
        \begin{array}{ccc}
          0 & 0 & 0 \\
          0 & 0 & 1 \\
          0 & 1 & 0
        \end{array}
      \right) &

      \left[\epsilon^{(4)}\right]_{ij} = \left(
        \begin{array}{ccc}
          1 & 0 & 0 \\
          0 & -1 & 0 \\
          0 & 0 & 0
        \end{array}
      \right) \\
      \\

      \left[\epsilon^{(5)}\right]_{ij} = \left(
        \begin{array}{ccc}
          0 & 1 & 0 \\
          1 & 0 & 0 \\
          0 & 0 & 0
        \end{array}
      \right) &

      \left[\epsilon^{(6)}\right]_{ij} = \left(
        \begin{array}{ccc}
          1 & 0 & 0 \\
          0 & 1 & 0 \\
          0 & 0 & 0
        \end{array}
      \right) \ .

   \end{array}
\end{equation}
From the above expressions it is easy to verify that the tensors
$\boldsymbol{\epsilon}^{(a)} \ , \ a = 1,....,6$ are linearly
independent and form an orthogonal basis. In our notation we have
labeled $4$ and $5$ the usual $+$ and $\times$ polarizations
respectively (the tensor polarization waveforms). The vector
polarizations ($s = \pm 1$) were labeled as $2$ and $3$, and finally
the two scalar modes ($s = 0$) have been denoted with the labels
$1$ (for the longitudinal scalar polarization) and $6$ (for the
transversal scalar mode).

Following \cite{burke1975,EW1975} we may notice that the space-time
described by the metric of Eq. (\ref{GW metric}) allows for the
following three Killing vectors
\begin{equation}
K^\rho_{(1)} = \delta^\rho_{x}, ~~~ K^\rho_{(2)} =
\delta^\rho_y, ~~~
K^\rho_{(3)} = \delta^\rho_z + \frac{\delta^\rho_t}{v} \ .
\label{killing 3}
\end{equation}
Generally speaking, in the weak field regime of a generic metric
theory of gravity, GWs can travel with $v \leq 1$. In that case the
linearized field equations are Klein-Gordon-type, and the resulting
group velocity is determined by the mass of the graviton, $m$.
Remembering that $v = \partial \omega/ \partial k$, and that the
dispersion relation is equal to $k = \sqrt{\omega^2 - m^2}$, we find
the following expression for $v$ in terms of $m$ and $\omega$
\begin{equation}
v(\omega) = \sqrt{1 - \left(\frac{m}{\omega}\right)^2} \ .
\label{v}
\end{equation}

Since the operational frequency band of LISA will be within the range
($10^{-4} - 1$ Hz), by assuming presently known upper-limits for the
mass of the graviton from Eq. (\ref{v}) we obtain the resulting values
for the group velocity of these waves.  After restoring physical units
and taking $m < 10^{-59} \ {\rm g}$ (the most stringent constraint to
date obtained by requiring the derived dynamical properties of a
galactic disk to be consistent with observations \cite{deAraujo2007}),
at $10^{-4}$ Hz we find a value for the group velocity $v$ whose
fractional difference, $\Delta_v$, from the speed of light is equal
to: $\Delta_v = |v - c|/c \simeq 10^{-15} $. A less stringent value
for the mass of the graviton equal to $m < 7.68 \times 10^{-55} \ {\rm
  g}$ (obtained from solar-system dynamics observations
\cite{Talmadge1988}), results into a $\Delta_v \simeq 10^{-8}$.  These
considerations imply that, no matter whether we are conservative or
not in our assumption about a likely upper limit for the graviton
mass, LISA will not be able to resolve the propagation speeds of the
different polarization components of the detected GWs by resolving the
time-separations of their imprints in the TDI combinations
\cite{AET1999}.  For this reason, from now on, we will assume $v = 1$
in natural units, and rewrite the third Killing vector
$\boldsymbol{K}_{(3)}$ as $K^\rho_{(3)} = \delta^\rho_z +
\delta^\rho_t$.

\begin{figure}
\centering
\includegraphics[width=6in]{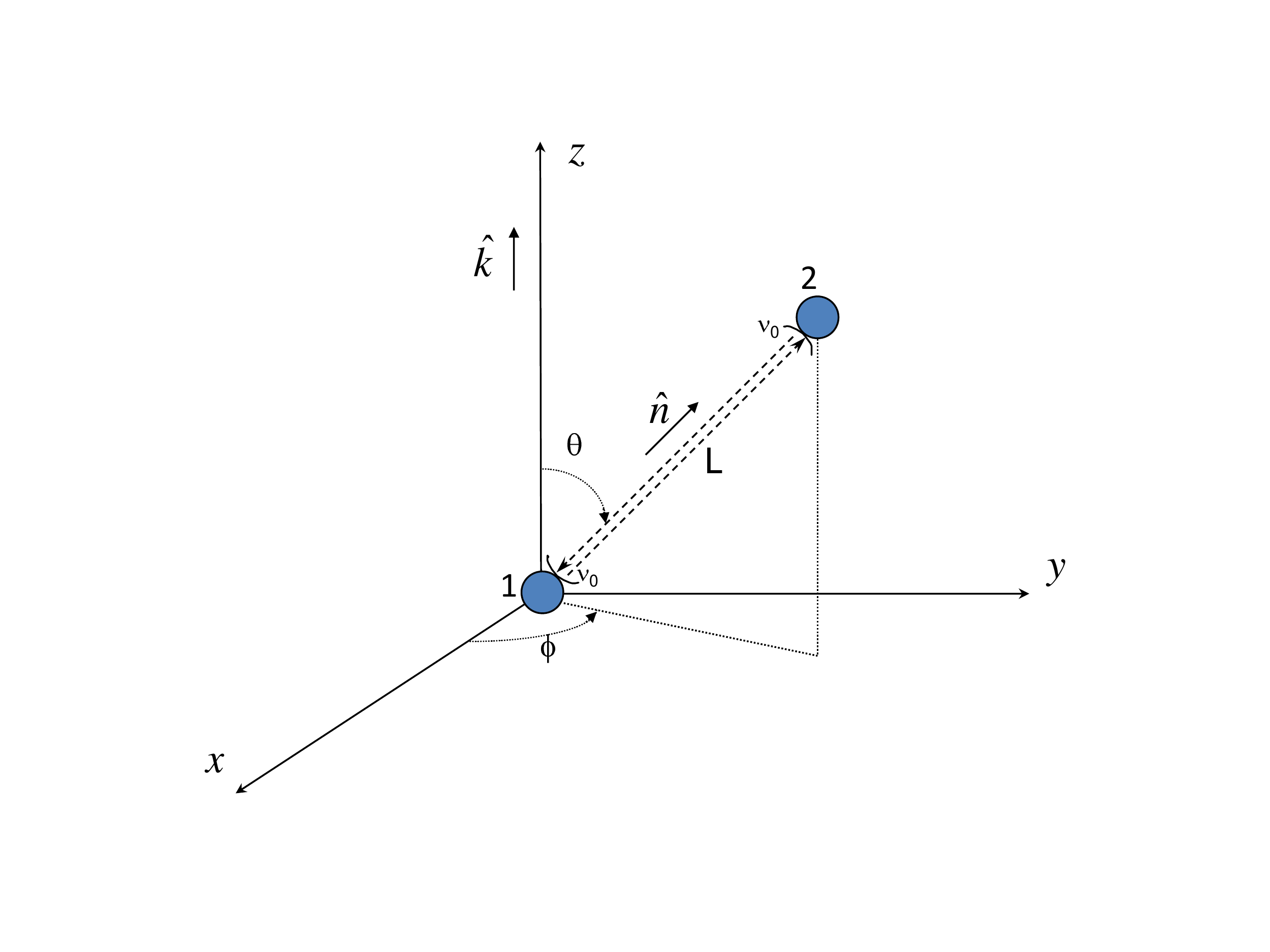}
\caption{A laser beam of nominal frequency $\nu_0$ is transmitted
  from spacecraft $1$ to spacecraft $2$ and simultaneously another
  beam from spacecraft $2$ is transmitted back to $1$. The
  gravitational wave train propagates along the $z$ direction, and the
  two polar angles ($\theta, \phi$) describe the direction of
  propagation of the laser beams relative to the wave. See text for a
  complete description.}
\label{Doppler}
\end{figure}

Let us now consider the unit vector $\hat n$ along the Doppler
link and oriented from spacecraft $1$ to $2$ (see Fig.
\ref{Doppler}). If we denote with ($\theta, \phi$) the usual polar
angles, $\hat n$ assumes the following familiar form
\begin{equation}
n^\rho = \sin\theta \cos \phi \ \delta^\rho_x + \sin \theta \sin \phi
\ \delta^\rho_y + \cos \theta \ \delta^\rho_z \ .
\label{unit v} 
\end{equation}
Since the most general Killing vector of our metric can be written as
a linear combination of the three Killing vectors above
\begin{equation}
K^\rho = a_1\delta^\rho_x + a_2 \delta^\rho_y + a_3(\delta^\rho_z +
\delta^\rho_t) \ ,
\label{general kil}
\end{equation}
with $a_1$, $a_2$, $a_3$ constants, by comparing Eq. (\ref{unit v}) and
Eq. (\ref{general kil}) we note that by taking $(a_1, a_2, a_3) = C \ \hat n$,
Eq. (\ref{general kil}) becomes
\begin{equation}\label{kill general}
K^\rho = C[n^\rho + {\hat k} \cdot {\hat n} 
\ \delta^\rho_t] \ ,
\end{equation}
where $C$ is an arbitrary constant, and $\hat k$ is the unit vector along
the direction of propagation of the wave (see Figure \ref{Doppler}).

If we now consider the 4-momentum of a photon transmitted by
spacecraft $1$, its analytic expression can be written in terms of the metric
perturbation at spacecraft $1$ in the following way \cite{burke1975,EW1975}
\begin{equation}
P_{\rho} = \nu_0 \left(-\delta_\rho^t + n_\rho +
  \frac{1}{2}h_{\rho \xi}n^\xi \right) \ ,
\end{equation}
where it is easy to see that the condition $P_{\rho}P^{\rho} = 0$ is
fulfilled to first order in the metric perturbation $h_{ij}$.  Since
$P_\rho K^{\rho} = constant$ along the photon world line
\cite{burke1975,EW1975}, we obtain the following relationship between
the frequency of the photon emitted at spacecraft $1$, $\nu_0$, and
that received at spacecraft $2$, $\nu^\prime$
\begin{equation}
P_\rho K^\rho = P_\rho^\prime K^{ \prime\rho},
\end{equation}
\begin{equation}
\nu_0 \left(-\delta_\rho^t + n_\rho +
  \frac{1}{2}h_{\rho\xi}n^\xi \right)(n^\rho + {\hat k} \cdot {\hat n}
\delta^\rho_t) 
=
\nu^\prime \left(-\delta_\rho^t + n^\prime_\rho + 
\frac{1}{2}h^\prime_{\rho\xi}n^{\prime\xi} \right)(n^{\prime\rho} +
  {\hat k} \cdot {\hat {n^\prime}}
\delta^\rho_t) \ .
\label{pconstant}
\end{equation}
If we now rewrite $n^{\prime\rho} = n^{\rho} + \delta n^\rho$, we may
notice that, to first order in $h_{ij}$, $n^\rho \delta n_\rho = 0$ because
$n_\rho^\prime n^\rho = n_\rho n^\rho = 1$. This result,
together with Eq.(\ref{pconstant}) above, allows us to derive the
following expression for the ratio of the two frequencies $\nu^\prime$
and $\nu_0$
\begin{equation}
\frac{\nu^\prime}{\nu_0} = \frac{1-{\hat k} \cdot {\hat n}+\frac{1}{2}n^\rho
  h_{\rho\xi}n^{\xi}}{1-{\hat k} \cdot {\hat n}+\frac{1}{2}n^\rho
  h^\prime_{\rho\xi}n^{\xi}} \ .
\end{equation}
Finally, expanding the right-hand-side of the last
equation to first order in $h_{\rho\xi}$ we get the resulting
expression for the one-way Doppler response, $y$
\begin{equation}
y  = (1 + \hat{k}\cdot\hat{n}) \ (\Psi - \Psi^\prime)
\end{equation}
where $y(t) \equiv (\nu^\prime (t) - \nu_0)/\nu_0$, and
$\Psi (t)$ is equal to
\begin{equation}
\Psi (t) \equiv \frac{n^i \ h_{ij} (t) \ n^j }{2[1 - (\hat{k}\cdot\hat{n})^2]}
\end{equation}
By explicitly showing the time dependence of the various terms,
$y (t)$ can be rewritten in the following form \cite{AET1999}
\begin{equation}
y(t) = (1 + \hat{k}\cdot\hat{n}) \ 
\left[\Psi(t - L) - \Psi(t - \hat{k}\cdot \hat{n} L) \right]
\label{oneway12}
\end{equation}
where $L$ is the separation between the two spacecraft. Note that, in
order to obtain the above expression, we have only assumed the
time-components of the metric perturbation, $h_{\rho t}$ to be equal
to zero \cite{Eardley1973}.

The expression for the one-way Doppler response measured on board
spacecraft $1$ at time $t$ can be obtained from Eq. (\ref{oneway12})
by changing ${\hat n} \to - {\hat n}$, and further delaying
the waveforms by $\hat k \cdot \hat n \ L$. The resulting
one-way Doppler response, $y^\prime (t)$, is equal to
\begin{equation}
y^\prime(t) = (1 - \hat{k}\cdot\hat{n}) \ 
\left[\Psi(t - (1 + \hat{k}\cdot\hat{n})L) - \Psi(t) \right]
\label{oneway21}
\end{equation}

Since the above expressions of the one-way Doppler responses are
identical {\it in form} to those valid for tensor waves
\cite{Wahlquist1987}, it follows that the various TDI combinations of
the LISA six inter-spacecraft one-way Doppler measurements will also
be identical {\it in form} to those derived in \cite{AET1999}. For these
reasons they will not be given here, and we refer the 
reader to \cite{TA1998,AET1999,TD2005} for more details.

\section{LISA Sensitivities}
\label{sensitivity}

In this section we will compute the LISA TDI sensitivities to vector
and scalar waves. Although the LISA sensitivities to tensor waves have
already been presented in an earlier publication \cite{ETA2000}, for
sake of comparison we will include them in the sensitivity plots
presented in this section.  We will specialize our calculations to the
equilateral-triangle configuration: ${\rm arm length \ 1} = {\rm arm
  length \ 2} = {\rm arm length \ 3} = L$ ($L \simeq 16.7$ light
seconds) since the LISA arm lengths will differ by at most a few
percent, and any corrections to our results will be to this level of
accuracy \cite{AET1999,ETA2000,TD2005}.

The LISA sensitivity to tensor GWs has been traditionally taken to be
equal to (on average over the sky and polarization states) the
strength of a sinusoidal GW required to achieve a signal-to-noise
ratio of $5$ in a one-year integration time, as a function of Fourier
frequency.  Although in the case of vector waves the average over the
polarization states can be performed by implementing the same
procedure described in \cite{AET1999,ETA2000} for tensor waves, in
general this can not be done for scalar waves.  This is because the
two scalar fields are mutually orthogonal (and independent), as one is
purely longitudinal and the other purely transverse to the direction
of propagation of the wave.

\begin{figure}
\centering
\includegraphics[width=6in, angle=0]{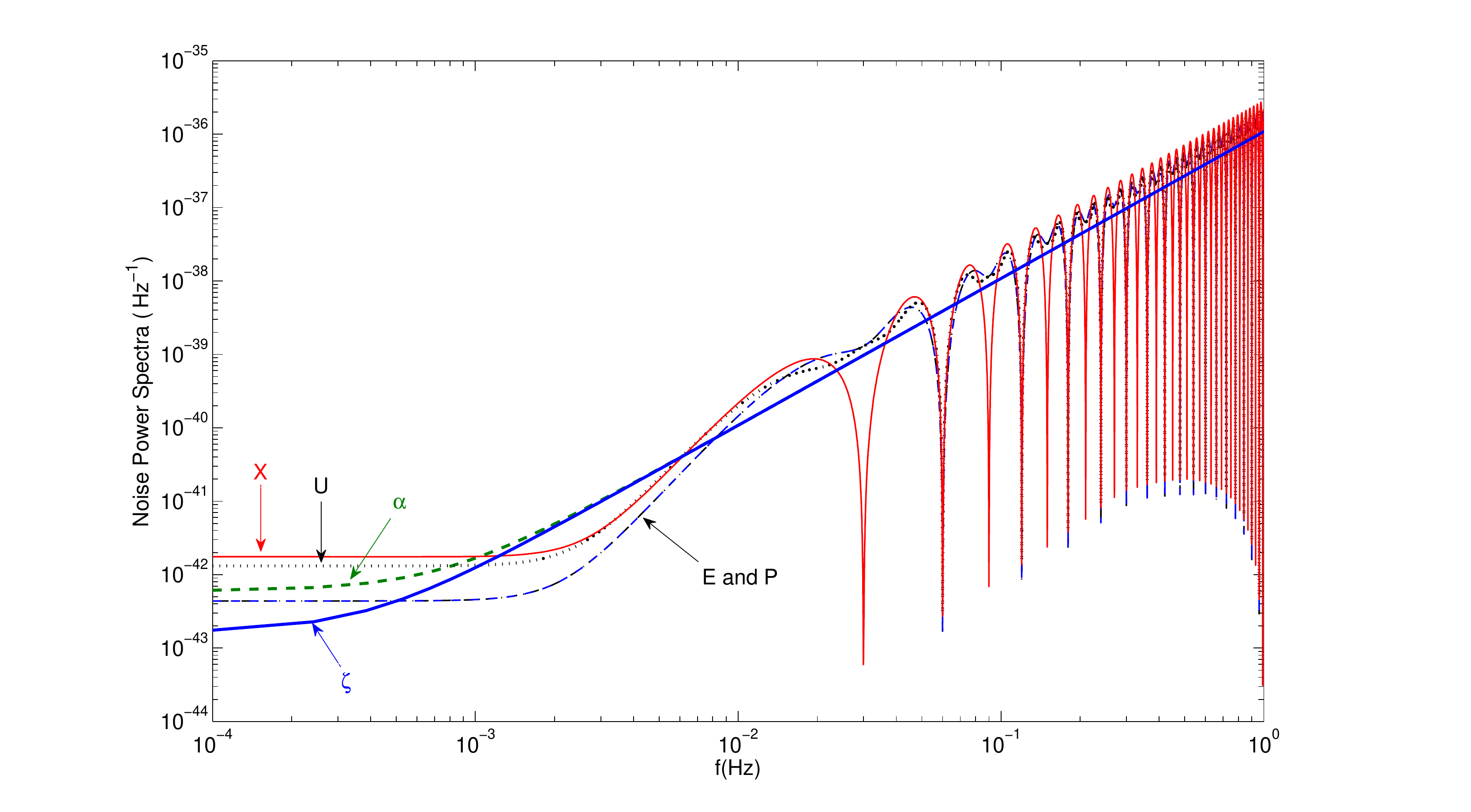}
\caption{Noise spectra in the $X$, $\alpha$, $\zeta$, $E$, $P$, and
  $U$ time-delay interferometric combinations accounting for the
  nominal proof-mass and optical path noises. The varying depths of
  the minima in the high-frequency ranges of $X$, $E$, and $P$ is an
  artifact of numerically calculating these functions at discrete
  frequencies}
\label{NoiseLISA}
\end{figure}

In order to compute the LISA sensitivities we will use the following
expressions for the power spectra of the noises affecting the $X$,
$\alpha$, $\zeta$, $E$, $P$, and $U$ combinations \cite{ETA2000} (see
Figure (\ref{NoiseLISA}))

\begin{eqnarray}
S_X (f) & = &
[8 \sin^2(4 \pi f L) + 32 \sin^2(2 \pi f L)] S^{pm}_y (f) +
16 \sin^2(2 \pi f L)\ S^{op}_y (f) 
\label{eq:10}
\\
S_{\alpha} (f) & = &   
[8 \sin^2(3 \pi f L) + 16 \sin^2(\pi f L)] S^{pm}_y (f) + 
6\ S^{op}_y (f) 
\label{eq:11}
\\
S_{\zeta} (f) & = &
24 \sin^2(\pi f L) \ S^{pm}_y (f) + 6\ S^{op}_y (f) 
\label{eq:12}
\\
S_E (f) & = &
[32 \sin^2(\pi f L) + 8 \sin^2(2 \pi f L)] S^{pm}_y (f) + 
[8 \sin^2(\pi f L) 
\nonumber
\\
& + & 8 \sin^2(2 \pi f L)] \ S^{op}_y (f) 
\label{eq:13}
\\
S_P (f) & = &
[8 \sin^2(2 \pi f L) + 32 \sin^2(\pi f L)] S^{pm}_y (f) + 
[8 \sin^2(2 \pi f L) 
\nonumber
\\
& + & 8 \sin^2(\pi f L)] \ S^{op}_y (f) 
\label{eq:14}
\\
S_U (f) & = &
[16 \sin^2(\pi f L) + 8 \sin^2(2 \pi f L) + 16 \sin^2(3 \pi f L)]
S^{pm}_y (f) 
\nonumber
\\
&+& 
[4 \sin^2(\pi f L) + 8 \sin^2(2 \pi f L) + 4 \sin^2(3 \pi f L)] \
S^{op}_y (f) \ ,
\label{eq:15}
\end{eqnarray}

where $S^{pm}_y (f) = 2.5 \times 10^{-48} \ [f /1 Hz]^{-2} \ {\rm
  Hz}^{-1}$ is the spectrum of the relative frequency fluctuations due
to each proof mass, and $S^{op}_y (f) = 1.8 \times 10^{-37}[f /1 Hz]^2
\ {\rm Hz}^{-1}$ is the spectrum of optical path (mainly shot and beam
pointing) noise. Both these noises can be regarded as the main
limiting noise sources for LISA \cite{PPA98,ETA2000}.

Gravitational wave sensitivity is the wave amplitude required to
achieve a given signal-to-noise ratio.  We calculate it in the
conventional way, requiring a signal-to-noise ratio of $5$ in a one
year integration time: $5 \ \sqrt{S_k(f) \ B}$/(root-mean-squared gravitational
wave response for data combination $k$), where $k$ is $X$, $\alpha$,
$\zeta$, $E$, $P$, $U$, and $S_k$ is the total noise power spectrum
for TDI combination $k$. The bandwidth, $B$, was taken to be equal to
one cycle/year (i.e. $3.17 \times 10^{-8}$ Hz).

We have assumed the vector-waves to be elliptically polarized and
monochromatic, with their wave functions, ($h^{(2)}$, $h^{(3)}$),
written in terms of a nominal wave amplitude, $H$, and the two
Poincar\'e parameters, ($\Phi, \Gamma)$, in the following way
\begin{eqnarray}
h^{(2)} (t) & = & H \ \sin(\Gamma) \ \sin(\omega t + \Phi) \ ,
\label{h2}
\\
h^{(3)} (t) & = & H \ \cos(\Gamma) \ \sin(\omega t) \ .
\label{h3}
\end{eqnarray}
For scalar signals instead, the two wave functions, ($h^{(1)}$,
$h^{(6)}$), have been treated as independent and we calculated the TDI
sensitivities to each of these two polarizations.  For both vector and
scalar signals we averaged over source direction by assuming uniform
distribution of the sources over the celestial sphere; in the case of
vector signals we also averaged over elliptical polarization states
uniformly distributed on the Poincar\'e sphere for each source
direction.  The averaging was done via Monte Carlo integration with
$4000$ source position/polarization state pairs per Fourier frequency
bin and $7000$ Fourier bins across the LISA ($10^{-4} - 1$) Hz band
\cite{AET1999,ETA2000}.

\begin{figure}
\centering
\includegraphics[width = 6in]{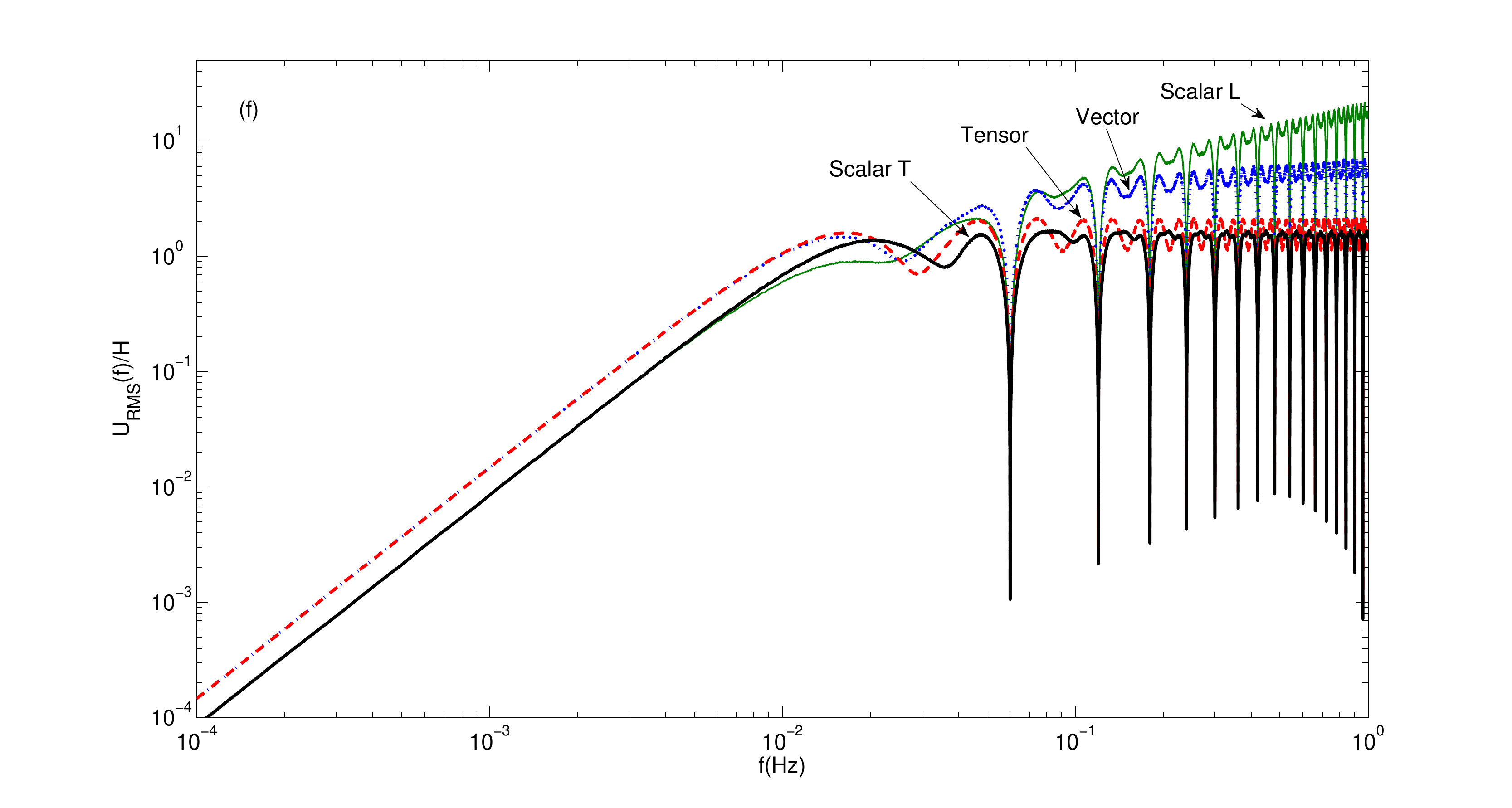}
\caption{Root-Mean-Square responses of the (a) X, (b) $\alpha$, (c)
  $\zeta$, (d) $E$, (e) $P$, and (f) $U$ TDI combinations to tensor
  ($s = \pm 2$), vector ($s = \pm 1$), and scalar ($s = 0$)
  gravitational waves. The former two have been treated as
  elliptically polarized waves, while the two waveforms characterizing
  the scalar signals have not as they are mutually orthogonal (and
  therefore independent). The sensitivities have been calculated by
  assuming an ensemble of sinusoidal signals uniformly distributed
  on the celestial sphere and (in the case of tensor and vector
  radiation) randomly polarized. In the high-part of the frequency
  band the r.m.s. values of the vector and scalar-longitudinal waves are
  noticeably larger than the tensor and scalar-transverse signals, and
  grow with the Fourier frequency.}
\label{RMS}
\end{figure}

Figure (\ref{RMS}) shows the root-mean-squared (r.m.s.) responses of
the TDI combinations (a) $X$ (unequal-arm Michelson), (b) $\alpha$
(Sagnac), (c) $\zeta$ (symmetrized Sagnac), (d) $E$ (monitor), (e) $P$
(beacon), and (f) $U$ (relay) to tensor (already derived in
\cite{AET1999} and shown here for comparison), vector and scalar
gravitational waves. In the high-part of the LISA frequency band we
may notice that the r.m.s.  responses to vector and
scalar-longitudinal waves are significantly larger than those to
tensor and scalar-transverse signals.  In particular, the
scalar-longitudinal r.m.s. response grows with the frequency at a much
faster rate than the others.

In order to physically understand this effect, let us compare a
one-way Doppler response (that measured on board spacecraft $1$, for
instance) to a ``pulse'' tensor wave against that due to a
scalar-longitudinal pulse wave. A tensor signal propagating
orthogonally to the light beam (direction for which the one-way
Doppler response can reach its maximum magnitude in this case) will
only interact with the light for the brief instance its wavefront
crosses the light beam. On the other hand, if a scalar-longitudinal
wave propagates along the direction between the two spacecraft (over
which the Doppler response will achieve its maximum in this case) the
frequency of the light will be affected by the gravitational wave for
the entire time $L$ it takes the wave to propagate from one spacecraft
to the other, resulting into an amplification of the frequency shift
when the wavelength of the wave is shorter than the inter-spacecraft
distance, $L$.  The above considerations become apparent by
considering in both cases the one-way Doppler response $y^\prime$ in
the Fourier domain.  For a tensor signal, the modulus-squared of the Fourier transform of Eq.
(\ref{oneway21}) with $\hat{k}\cdot\hat{n} \rightarrow 0$ becomes
equal to
\begin{equation}
|{\widetilde {y^\prime}} (f)_{{\hat k} \cdot {\hat n} \rightarrow 0}|^2
=  \sin^2(\pi f L) \ |{\widetilde h} (f)|^2 \ ,
\end{equation}
with ${\widetilde h} \equiv {\widetilde h}^{(4)} \cos(2 \phi) +
{\widetilde h}^{(5)} \sin(2 \phi)$. In the case of a
scalar-longitudinal signal instead, it is easy to show that, in the
limit of ${\hat k} \cdot {\hat n} \rightarrow -1$ the modulus-squared
of the Fourier transform of Eq. (\ref{oneway21}) becomes equal to
\begin{equation}
|{\widetilde {y^\prime}} (f)_{{\hat k} \cdot {\hat n} \rightarrow -1}|^2
=  (\pi f L)^2 \ |{\widetilde h}^{(1)} (f)|^2 \ .
\label{amplification}
\end{equation}
From the above two expressions we may conclude that, at frequencies
larger than the inverse of the one-way-light-time and for tensor and
scalar-longitudinal waves of comparable amplitudes, the maximum
one-way Doppler response to a scalar-longitudinal signal will be
larger than the corresponding maximum response to a tensor wave by
roughly a factor of $\pi f L$. This example implies that the r.m.s. of
the TDI responses to scalar-longitudinal signals will be larger than
those to tensor waves in the high-frequency region of the LISA band.
Similar considerations can be made for understanding the differences
between the r.m.s. responses to tensor and vector waves.

In the low-frequency limit (for $f < 5 \times 10^{-3} $ Hz)
the tensor and vector r.m.s. responses coincide, while those for the
two scalars also coincide with each other but are smaller by a factor
of about $2$ than those for tensor and vector waves. This is because
the two scalar waves, being mutually orthogonal, have been treated as
independent rather than elliptically polarized like the vector and
tensor waves.

\begin{figure}
\centering
\includegraphics[width = 6in]{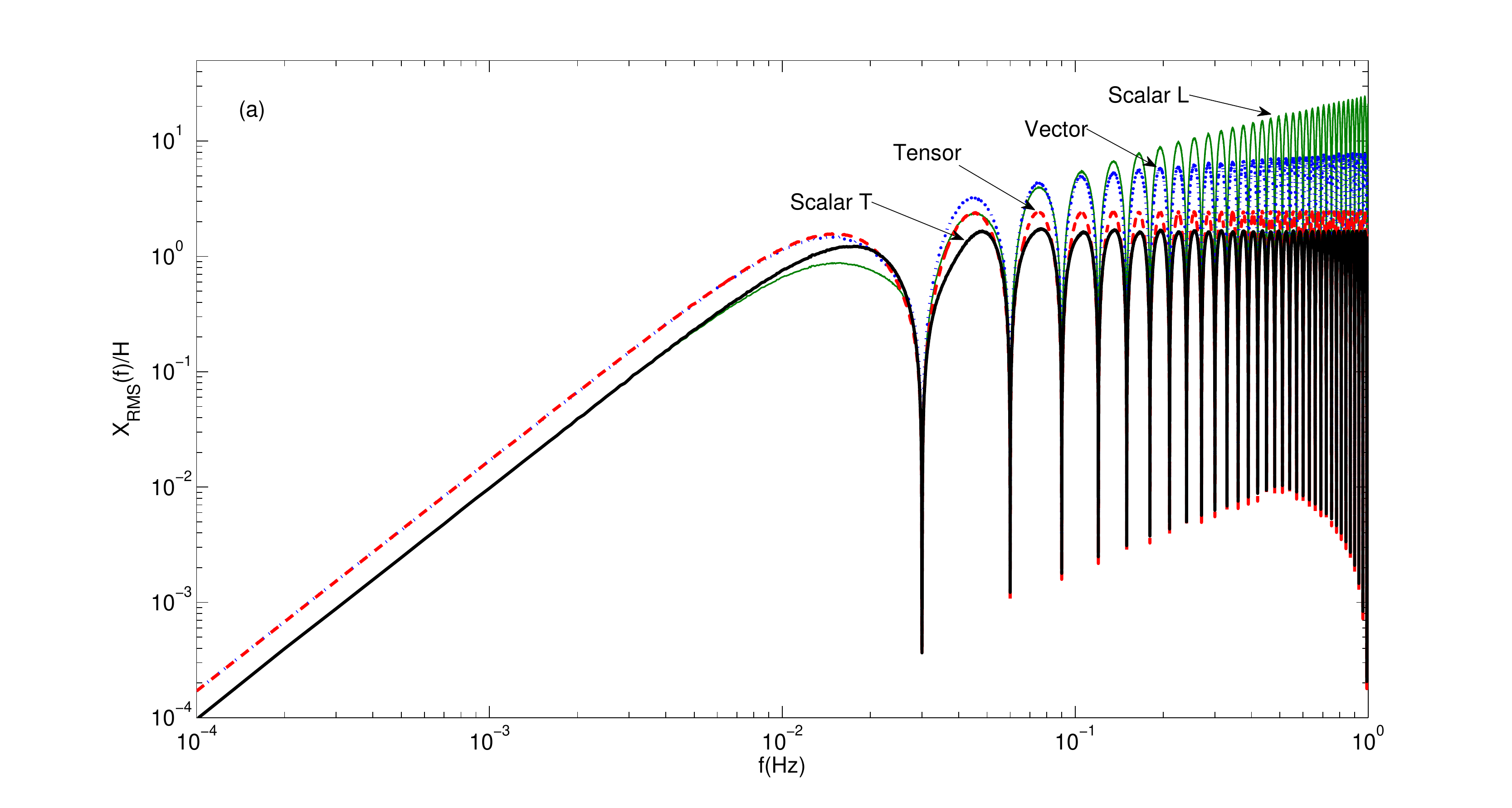}
\end{figure}

\begin{figure}
\centering
\includegraphics[width = 6in]{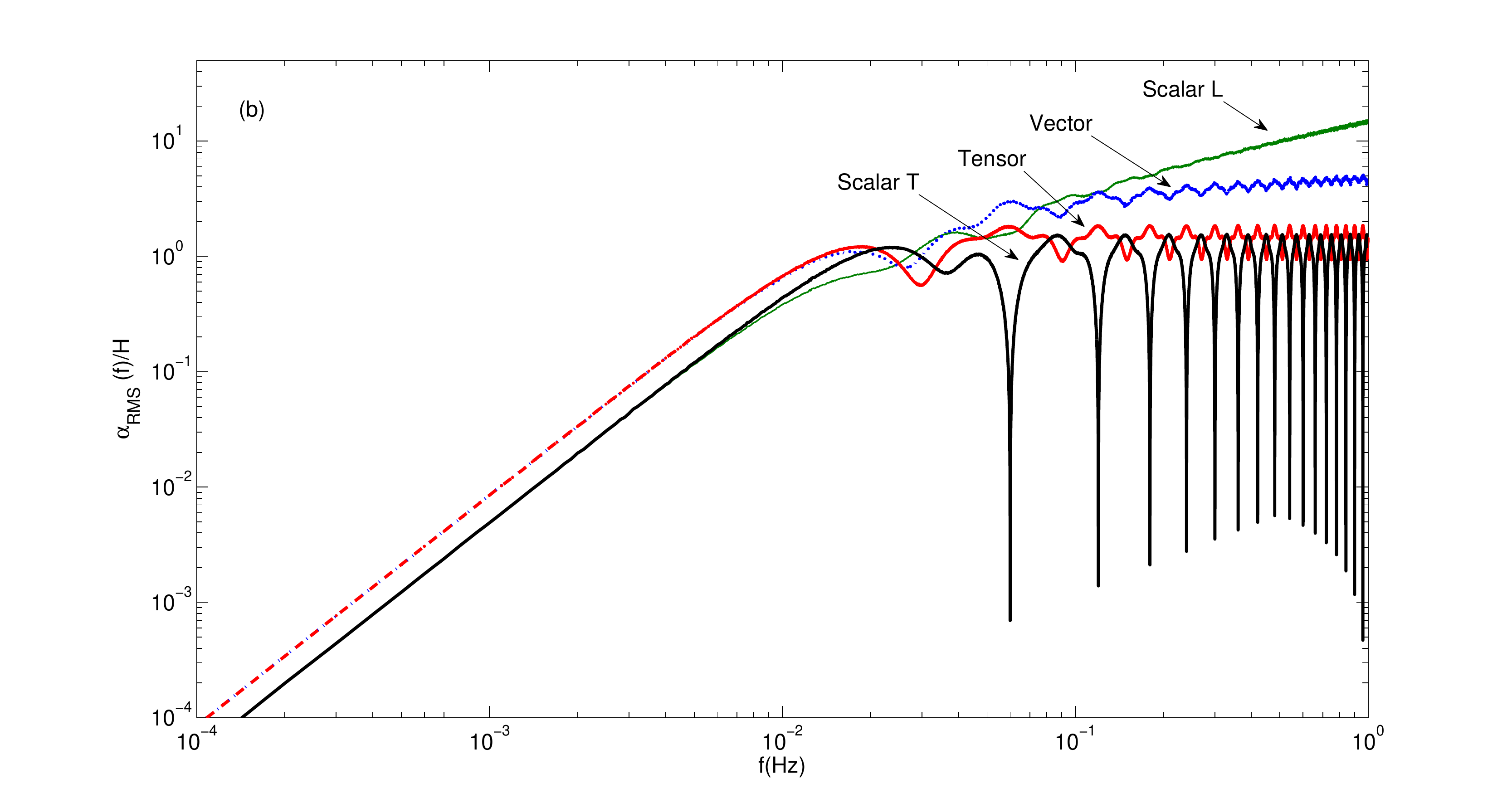}
\end{figure}

\begin{figure}
\centering
\includegraphics[width = 6in]{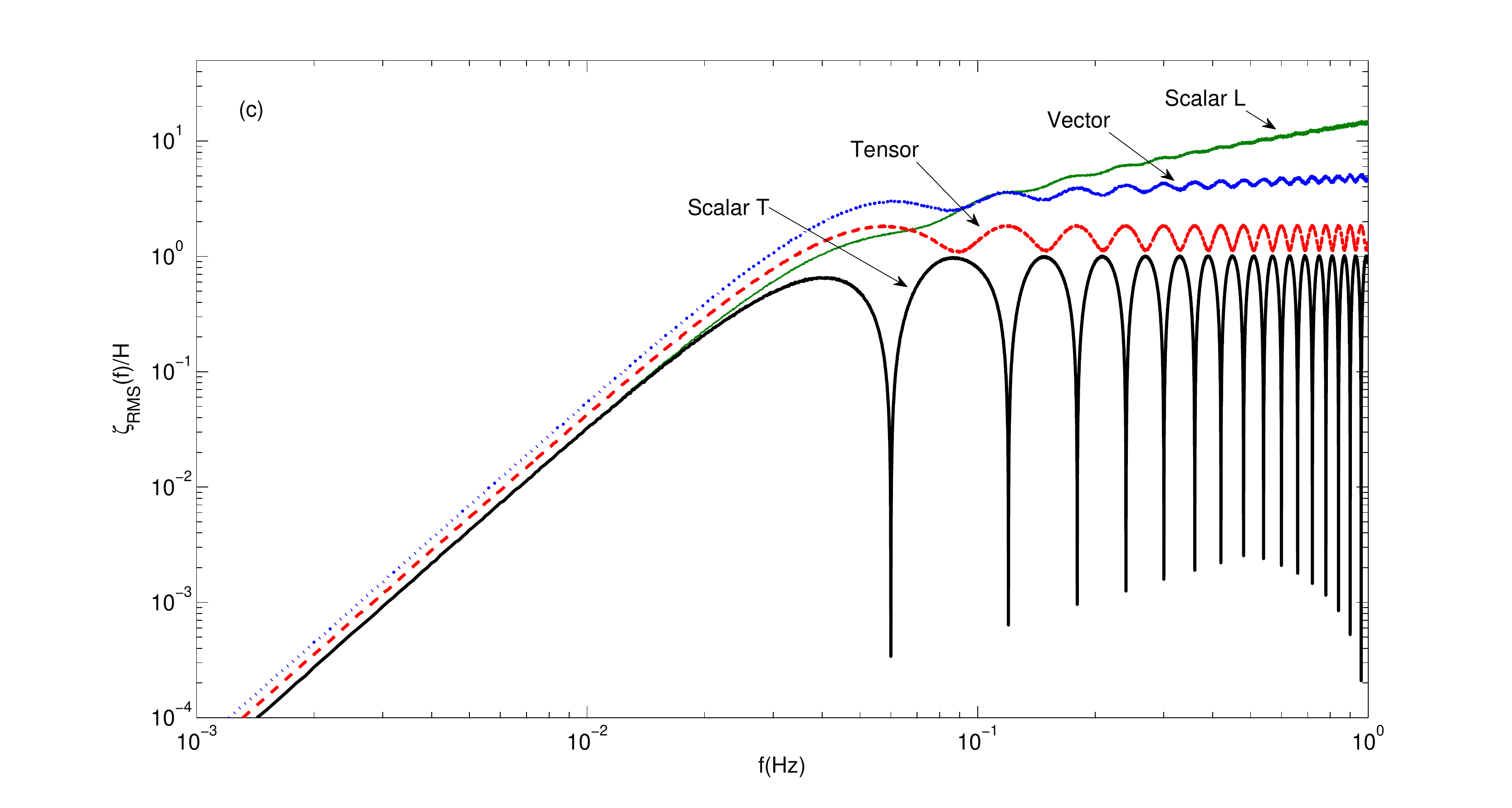}
\end{figure}

\begin{figure}
\centering
\includegraphics[width = 6in]{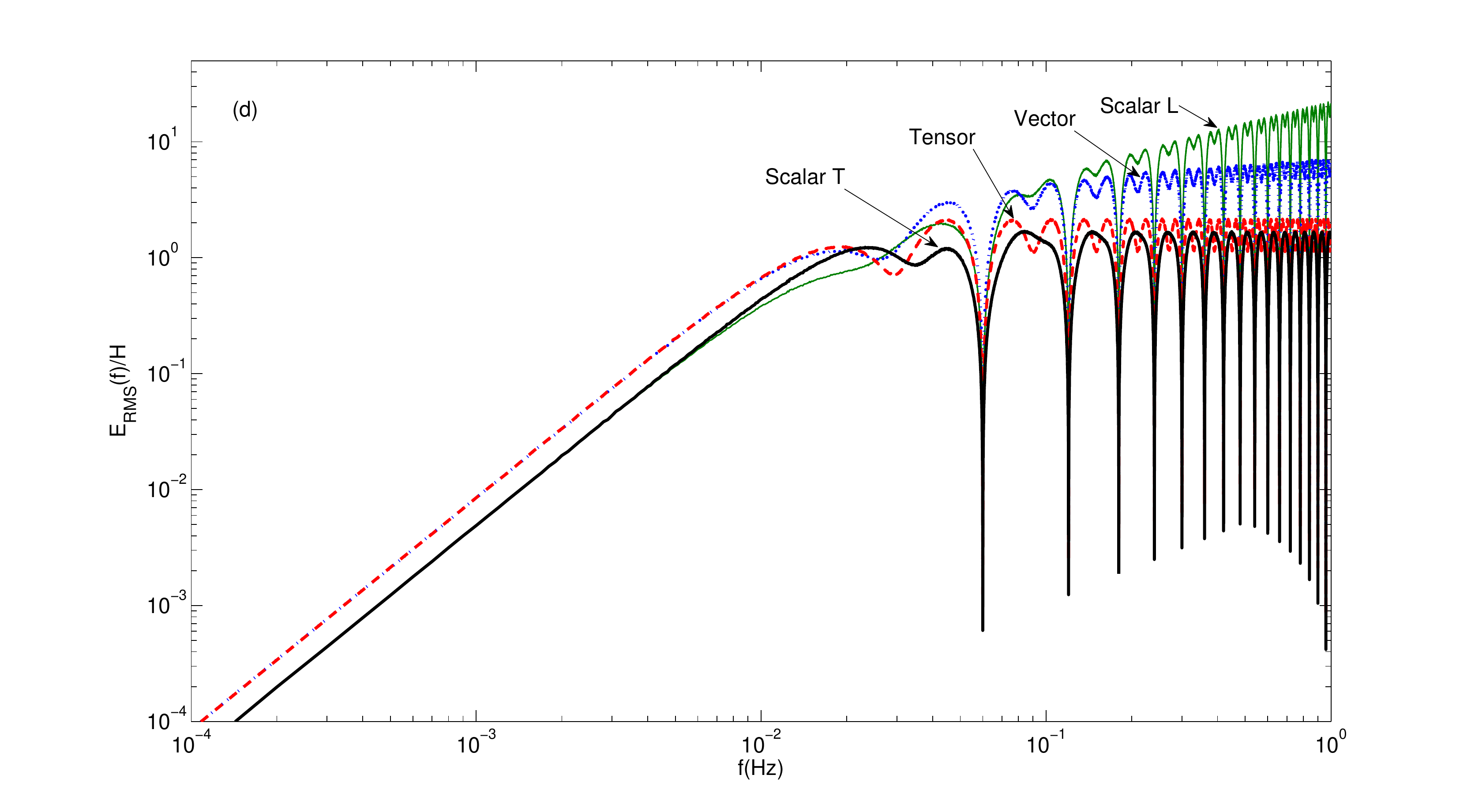}
\end{figure}

\begin{figure}
\centering
\includegraphics[width = 6in]{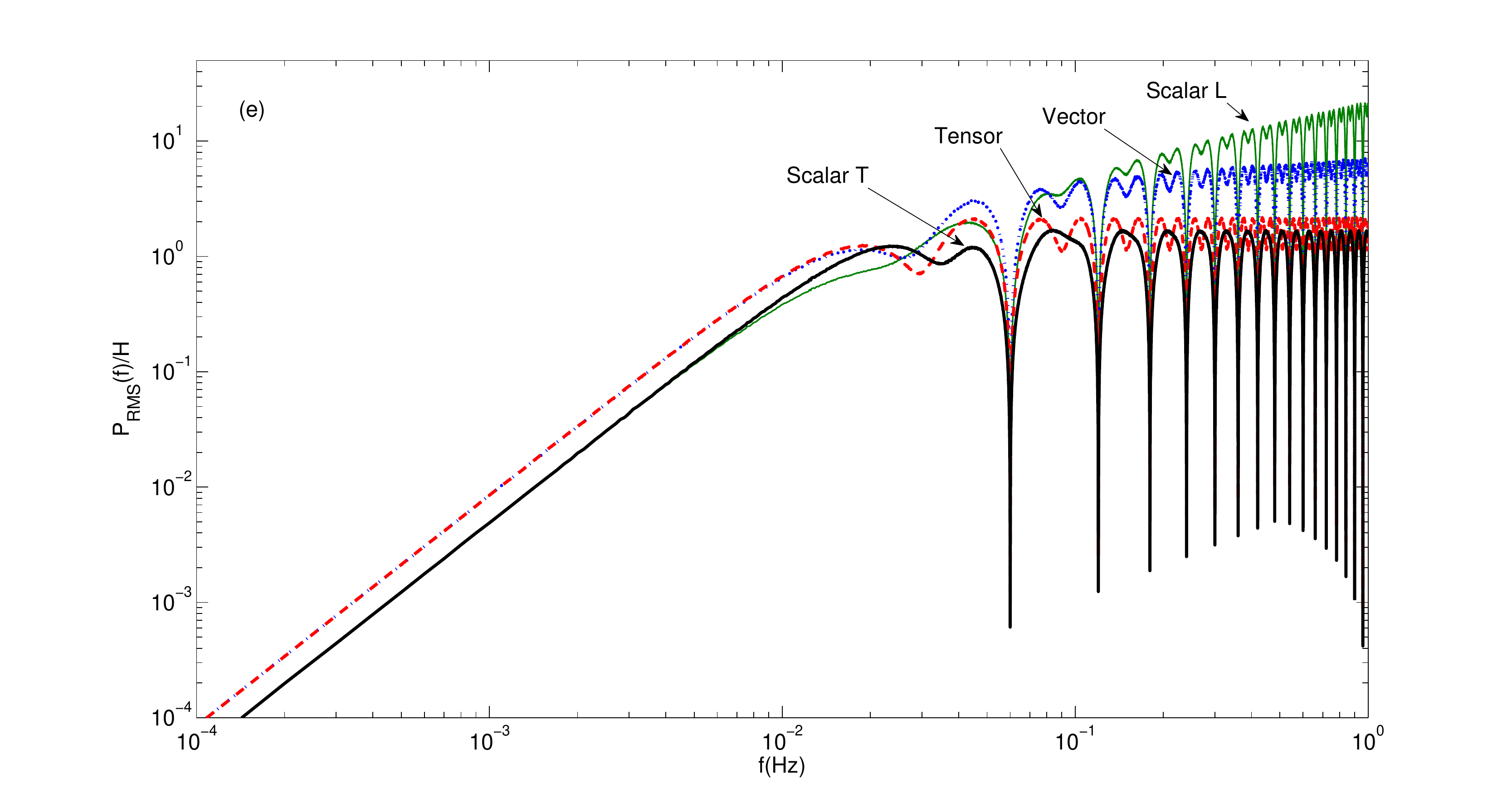}
\end{figure}

\begin{figure}
\centering
\includegraphics[width = 6in]{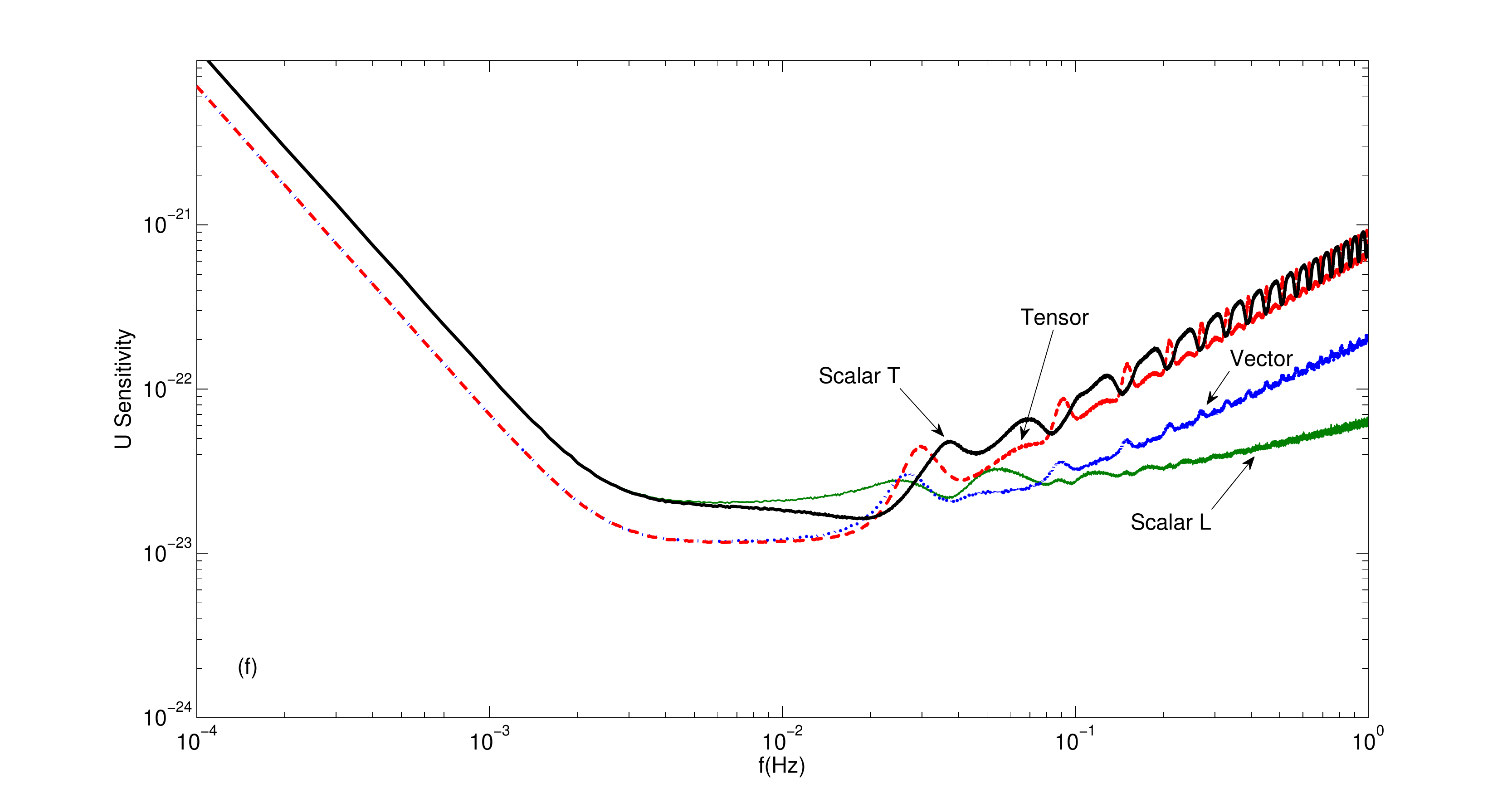}
\caption{Sensitivities of the (a) X, (b) $\alpha$, (c)
  $\zeta$, (d) $E$, (e) $P$, and (f) $U$ TDI combinations to
  gravitational waves with tensor ($s = \pm 2$), vector ($s = \pm 1$),
  and scalar ($s = 0$) components. Consistently with Figure
  (\ref{RMS}) we may notice how more sensitive the various TDI
  combinations are to vector and scalar-longitudinal signals than to
  tensor and scalar-transverse waves in the high-frequency part of the
  LISA frequency band. See text for more details.}
\label{Sensitivities}
\end{figure}

In Figure (\ref{Sensitivities}) we then plot the corresponding
sensitivities of the TDI combinations to the tensor, vector and the
two scalar polarization components. The characteristic behavior of the
r.m.s.  responses to scalar-longitudinal and vector signals shown in
figure (\ref{RMS}) folds into the plots presented here. At
high-frequencies the sensitivity to scalar-longitudinal waves is
significantly better than that to tensor, vector, and
scalar-transverse waves. At $1$ Hz, for instance, the sensitivity of
scalar-longitudinal signals is about a factor of $fL \simeq 17$ better than that
to tensor waves, while it is only a factor of $3$ better than the
sensitivity to vector signals.

Another interesting feature noticeable in Figure (\ref{Sensitivities}) is the lack
of sensitivity of the Sagnac combinations (b) $\alpha$ and (c) $\zeta$ to
scalar-transverse waves at frequencies equal to integer-multiples of
the inverse of the one-way-light-time. We have verified this result
analytically and found that indeed, at these frequencies, the Sagnac
responses are identically equal to zero independently of the
direction of propagation of the signal.

\begin{figure}
\centering
\includegraphics[width = 6in]{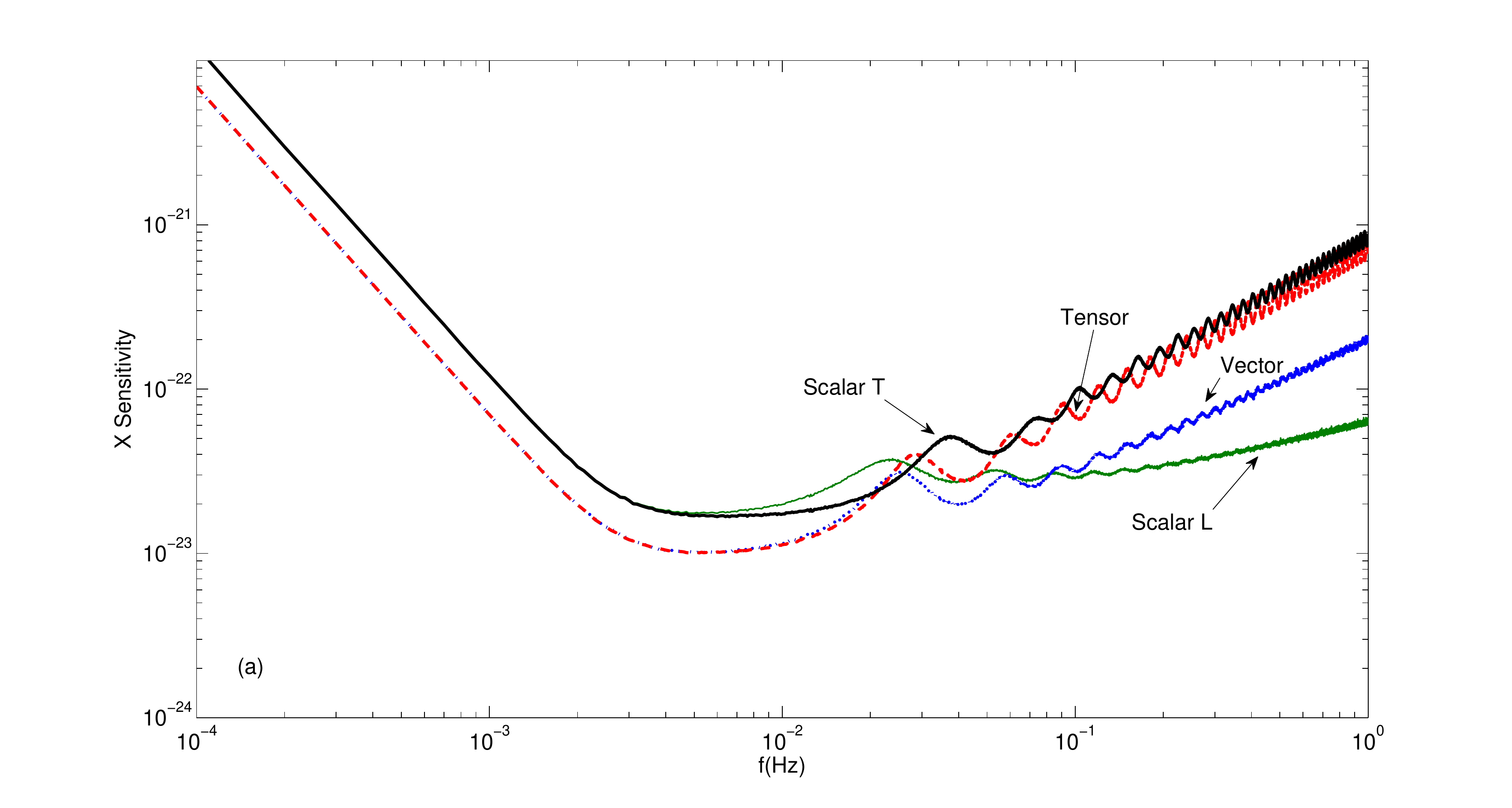}
\end{figure}
\begin{figure}
\centering
\includegraphics[width = 6in]{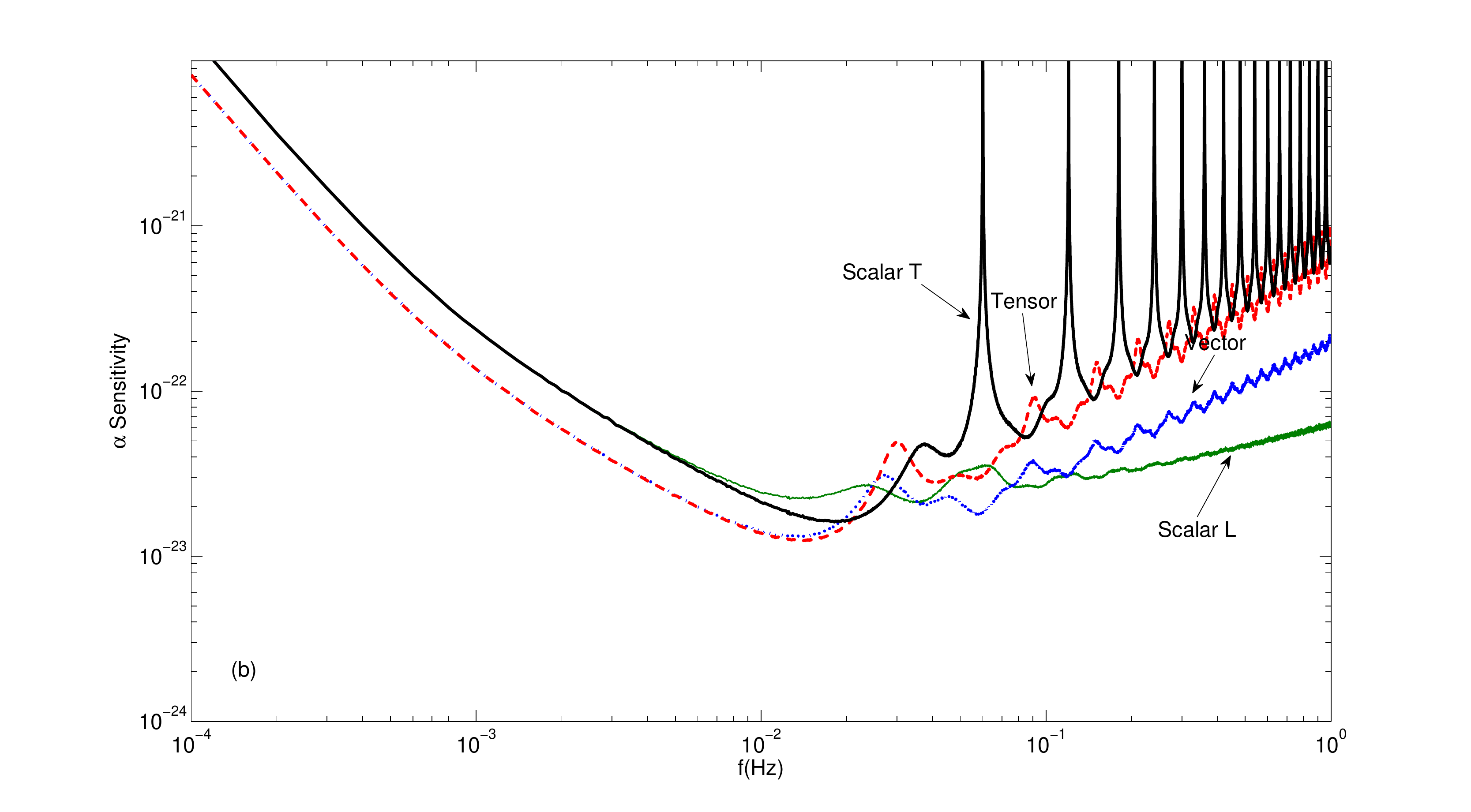}
\end{figure}
\begin{figure}
\centering
\includegraphics[width = 6in]{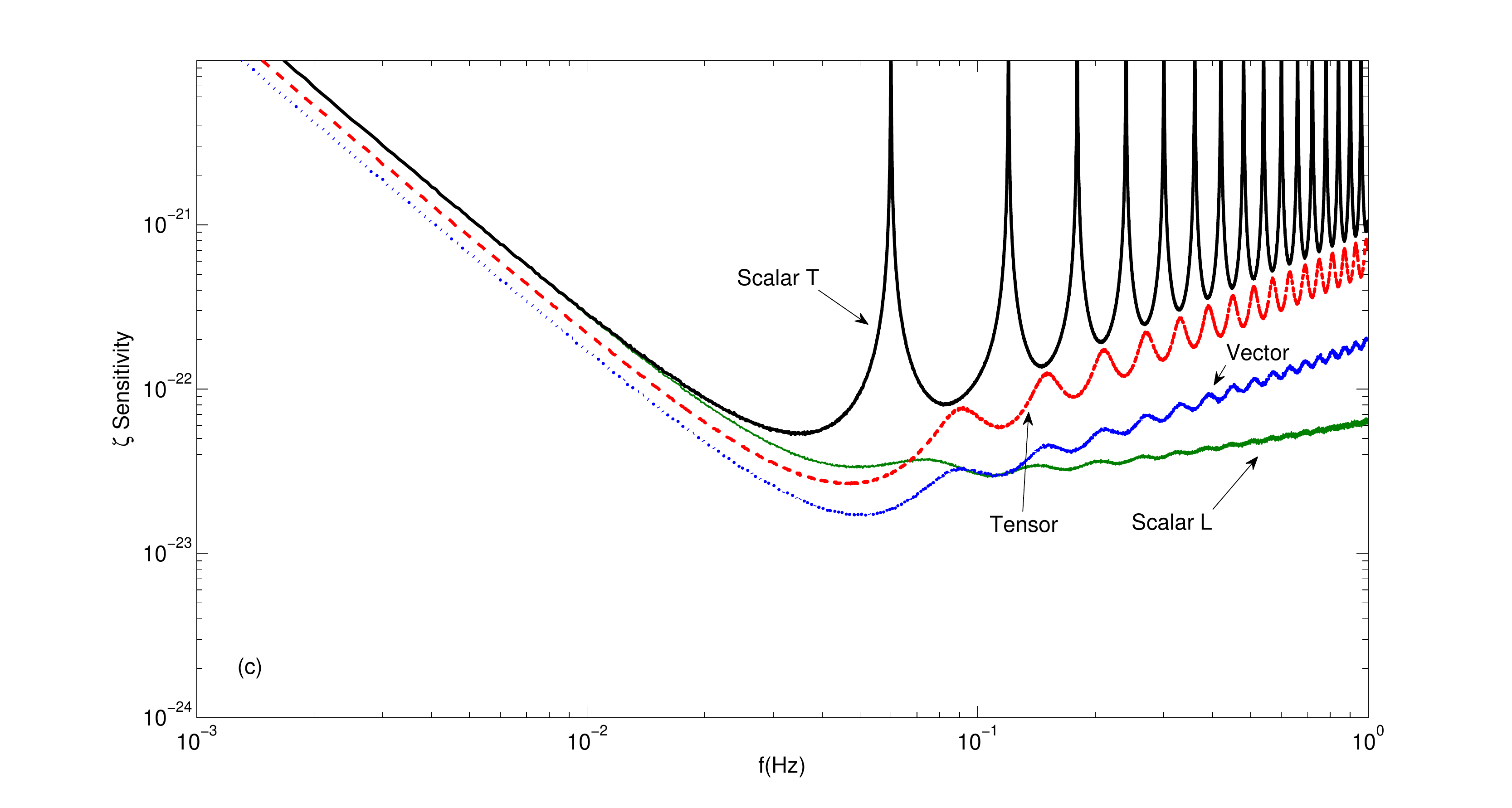}
\end{figure}
\begin{figure}
\centering
\includegraphics[width = 6in]{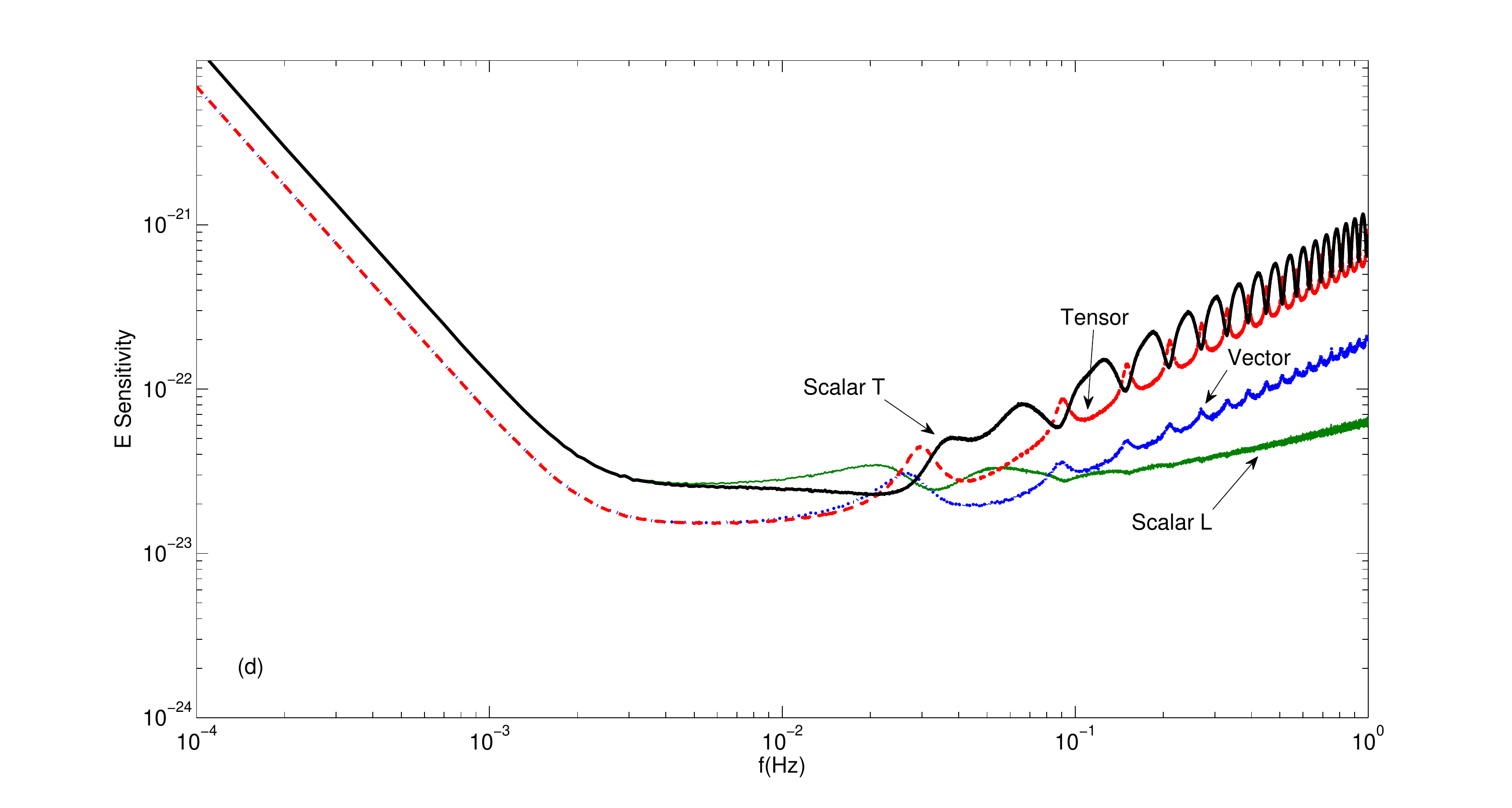}
\end{figure}
\begin{figure}
\centering
\includegraphics[width = 6in]{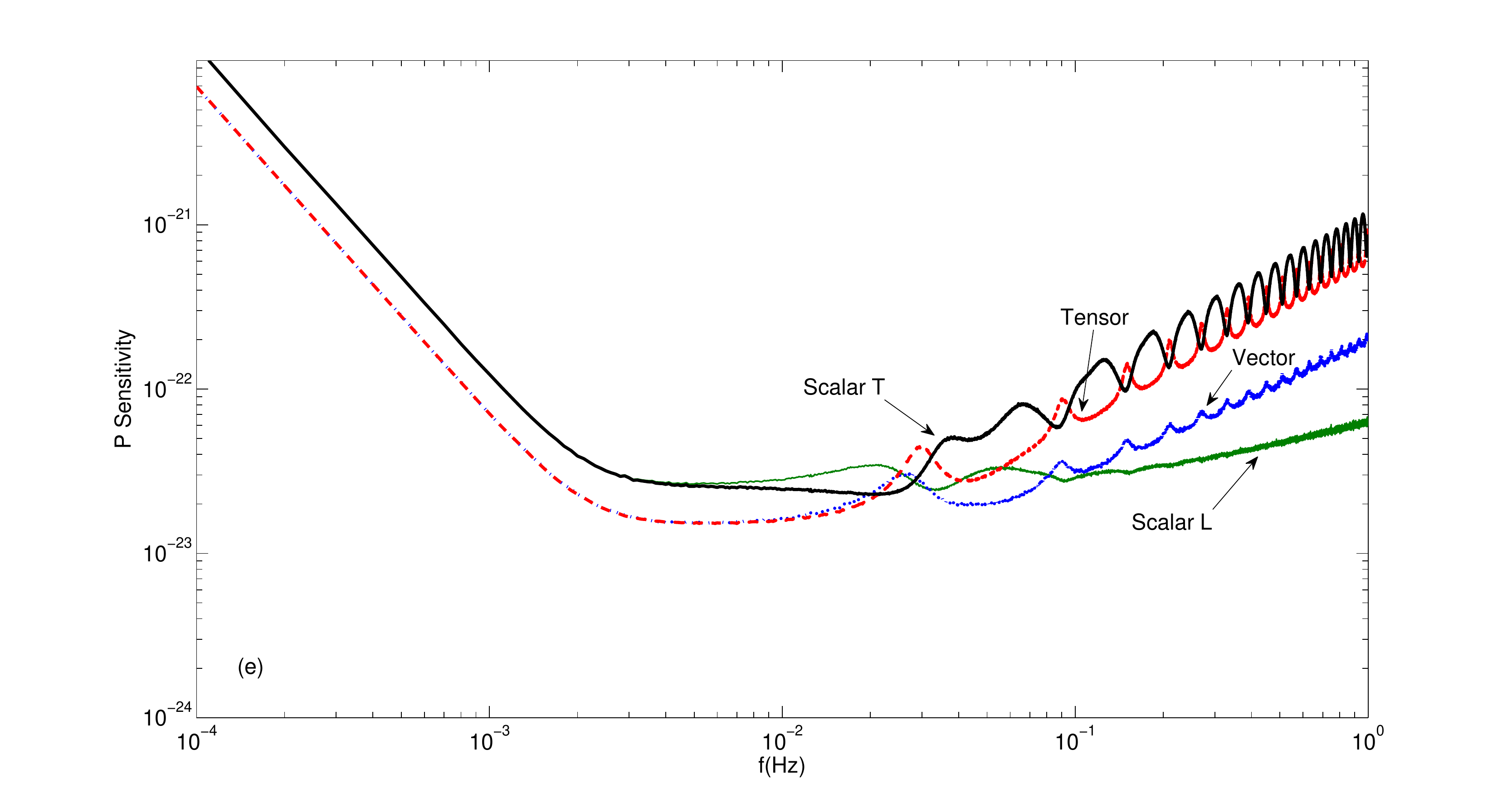}
\end{figure}

\section{Summary and Conclusions}
\label{Conclusions}

The main results of our work have been that (i) LISA is more sensitive
to scalar-longitudinal and vector signals than to tensor waves in the
high-part of its frequency band, and (ii) at low-frequencies its
sensitivities to tensor and vector signals are equal and somewhat
better than those to scalar waves. We have also found that the LISA
TDI Doppler responses to scalar-longitudinal waves propagating along
any of the three LISA arms will experience an amplification proportional
to the arm-length.  These results, together with the LISA capability
for constructing three independent TDI combinations in the high part
of its observational frequency band, should provide LISA with the
capability for assessing the polarization of the waves it will detect.
This will be the topic of our forthcoming investigation.

\section*{Acknowledgments}

We would like to thank Odylio Aguiar and Mario Novello for their kind
hospitality at their institutions while this research was first
formulated, and for financial support.  We also thank Frank B.
Estabrook and John W. Armstrong for their constant encouragement
during the development of this work. This research was performed at
the Jet Propulsion Laboratory, California Institute of Technology,
under contract with the National Aeronautics and Space
Administration.(c) 2008 California Institute of Technology.
Government sponsorship acknowledged.

\end{document}